%% file: main.tex
\documentclass[12pt,a4paper,final]{iopart}

\usepackage[latin1]{inputenc}
\usepackage{mathptmx}
\usepackage{graphicx}
\usepackage{iopams}
\usepackage{url}

\newcommand*{\eg}{\emph{e.g.}}

\newcommand*{\ie}{\emph{i.e.}}

\begin{document}

\title{Nonlinear physics of the ionosphere and LOIS/LOFAR}

\author{Bo Thid\'e\thanks{Also at LOIS Space Centre, MSI, V\"axj\"o
University, SE-351\,95 V\"axj\"o, Sweden}}

\address{Swedish Institute of Space Physics,
P.\,O. Box 537, SE-751\,21 Uppsala, Sweden}

\ead{bt@irfu.se}

\input{abstract}
\input{introduction}
\input{newdiagnostics}

\input{acknowledgements}

\input{references}
\end{document}

%% file: abstract.tex
\begin{abstract}

The ionosphere is the only large-scale plasma laboratory without walls
that we have direct access to.  Here we can study, both in situ and
from the ground, basic small- and large-scale processes and fundamental
physical principles that control planet Earth's interaction with its
space environment.  From results obtained in systematic, repeatable
experiments, where we can vary the stimulus and observe its response in
a controlled, laboratory-like manner, we can draw conclusions on similar
physical processes occurring naturally in the Earth's plasma environment
as well as in parts of the plasma universe that are not easily accessible to
direct probing.

Of particular interest is electromagnetic turbulence excited in the
ionosphere by beams of particles (photons, electrons) and its
manifestation in terms of secondary radiation (electrostatic and
electromagnetic waves), structure formation (solitons, cavitons,
alfveons, hybrons, striations), and the associated exchange of energy,
linear momentum, and angular momentum.

The primarily astrophysics-oriented, distributed radio telescope Low
Frequency Array (LOFAR) currently under construction in the Netherlands,
Germany, and France, will operate in a frequency range (10--240~MHz),
close to fundamental ionospheric plasma resonance/cut-off frequencies,
with a sensitivity that is orders of magnitude higher than any radio (or
radar) facility used so far.  The LOFAR Outrigger in Scandinavia (LOIS)
radio and radar facility, with one station in V\"axj\"o in southern Sweden
and three more planned in the same area (Ronneby, Kalmar, Lund) plus one
near Poznan in Poland, supplements LOFAR with optimised Earth and space
observing extensions.  For this purpose LOIS will operate in the same
frequency range as LOFAR (but extended on the low-frequency side) and
will augment the observation capability to enable direct radio imaging of
plasma vorticity.

\end{abstract}

%% file: introduction.tex
\section{Introduction}

The understanding of the dynamic interaction and structure formation
phenomena occurring in the Earth's upper atmosphere and ionosphere is
of ever-increasing importance.  The magnetised plasma of the ionosphere
is a non-linear medium in which different kinds of turbulence are easily
produced by a source of free energy in the form of natural and man-made
perturbations.  Of particular interest is the transfer of energy, momentum
and angular momentum from the source, \eg, powerful electromagnetic waves
emitted by radio transmitters, to the turbulent structures, as well as
the conversion of electrostatic turbulence into electromagnetic waves and
energised electrons.  Traditionally, plasma structures of various spatial
and temporal scales in the ionospheric and magnetospheric plasma have been
investigated by means of a variety of radio-physical methods using both in
situ and remote diagnostic techniques.  These investigations improve our
understanding of the nature of the plasma turbulence at various altitudes
in the Earth's atmosphere and ionosphere and can be extrapolated to the
study of non-linear structures in space and laboratory plasma.  However,
studies under natural conditions are associated with severe disadvantages
and limitations because many external and irregular factors that exert an
influence on the results are often not known.  The use of powerful radio
waves to systematically create ionospheric plasma structures (artificial
ionospheric turbulence, AIT) with controllable and repeatable properties
provides an advanced approach for solving this and related problems as
well as possibilities to develop new diagnostics.

The possibility that the ionosphere could be modified by powerful radio
waves was first noted by \emph{Ginzburg and Gurevich}
\cite{Ginzburg&Gurevich:SPU:1960}.  The early
theoretical work concentrated on the heating caused by the powerful radio
wave, but later the emphasis gradually changed to plasma instabilities,
turbulence, and plasma structuring.  The first ionospheric modification
facility was built in 1961 near Moscow, Russia, followed by facilities
in Colorado, in Puerto Rico, at several additional sites in the former
Soviet Union, in Norway, and in Alaska.  AIT is currently being studied
at research facilities located at middle (Sura, Russia) and high (EISCAT,
Norway; HAARP and HIPAS, Alaska, USA) latitudes.  In addition, a low
latitude facility (Arecibo, Puerto Rico, USA) was active until 1998 and
is now being rebuilt.  Under construction in Europe is the huge LOFAR
(Low Frequency Array), financed by the Dutch government.  This 10--240 MHz
radio telescope is of a new digital type which ensures maximum flexibility
and cost effectiveness, allowing it to become the world's largest and most
efficient instrument for low-frequency radio studies of space.  LOFAR is
being supplemented by a likewise digital and cost effective infrastructure
in Southern Sweden called LOIS (LOFAR Outrigger in Scandinavia).

This field of physics, encompassing the experimental and theoretical
study of the complex interplay between electromagnetic waves and plasma
turbulence, is important not only within the framework of
ionospheric modification but in plasma physics in general.  Particular
applications include circumsolar and circumstellar plasma, where
escaping electromagnetic waves carry essential diagnostic information on
the systems from which they emanate, and research on the physics and
technological development of nuclear fusion reactors for energy
production, where plasma instabilities generated by powerful
electromagnetic laser pump beams have been a major concern.  It is therefore
important that maximum information is extracted from the radio emissions
associated with the turbulence.

%% file: newdiagnostics.tex
\section*{New, information-rich radio diagnostics}

Classical electrodynamics exhibits a rich set of symmetries
\cite{Ribaric&Sustersic:Book:1990} and for each Lie symmetry there exists
an associated conserved quantity
\cite{Noether:NGWG:1918,Ibragimov:JMAA:2007}.  Commonly utilised
conserved EM quantities are the energy and linear momentum, where the
underlying symmetries, under Poincar\'e transformations, are homogeneity
in time and homogeneity in space, respectively.  An everyday physics
manifestation of the linear momentum conservation is the (translational)
Doppler effect.

Another conserved EM quantity is the angular momentum, introduced into
the theory already a century ago \cite{Poynting:PRSL:1909} and
demonstrated experimentally in 1936 \cite{Beth:PR:1936}.  This property
of EM beams has come to the fore during the past couple of decades in
optics \cite{Allen:JOB:2002} as well as in atomic and molecular physics
\cite{Cohen-Tannoudji:RMP:1998}, but has not yet been utilised to any
significant degree in radio physics
\cite{Krishnamurthy&al:Asilomar:2004} or its applications such as radio
astronomy \cite{Harwit:APJ:2003}.  

In order to utilise a radio antenna array for orbital angular momentum
(OAM), the individual antennas must be able to sense the full 3D vectors
of the EM field over an area that is large enough to intersect a
substantial fraction of the radio beam.  In the general case, one must
use vector antennas, \eg, tripoles
\cite{Compton:IEEETAP:1981,Carozzi&al:PRE:2000}; conventional
information-wasting crossed dipoles will not be optimum for radio beam
axes far from perpendicular to the plane spanned by the two antenna
elements.  The use of tri-channel digital radio systems that operate
with enough amplitude resolution at high enough sampling rates and that are
connected directly to the individual antenna elements, makes it possible
to sample coherently the instantaneous 3D radio field vectors up to
several GHz.  This enables the processing of EM field vectors, including
OAM encoding and decoding of radio beams, with high precision and speed
under full software control.  This is in contrast to optics where
detectors are still incoherent, capable of measuring second (and
sometimes higher) order field quantities only and not the field vectors
themselves.

Simple one-dimensional sensing antennas are what is typically used
for picking up radio and TV broadcasts on domestic radio and TV
sets but are also used in more demanding situations.  Two-dimensional
sensing of the 3D vector field (\eg, crossed dipole antennas) is used in
many modern radio telescopes, including LOFAR.  Traditional sensing of
the radio field in two dimensions is also going to be used for the next
generation antennas of the planned ``3D'' European Incoherent Scatter
(EISCAT) ionospheric radar facility;\footnote{\url{www.eiscat.se}} here
``3D'' does not refer to the way the EM fields of the radar are sensed
but how the radar signals, once they have been sensed with a standard 2D
``information wasting'' technique, will be used in attempts to estimate the
3D plasma dynamics in the ionosphere.  The Scandinavian supplement to
LOFAR, LOIS, is the first space physics radio/radar or radio astronomy
facility to utilise the entire 3D vector information embedded in the EM
fields of the radio signals.  Future big Earth-based radio astronomy
multi-antenna telescopes such as the Square Kilometre Array
(SKA),\footnote{\url{www.skatelescope.org}} and the Long Wavelength
Array (LWA),\footnote{\url{lwa.nrl.navy.mil}} are expected to benefit
significantly from using LOIS-type vector-sensing radio technology.
This is even more true for space-based radio infrastructures such as the
proposed Lunar Infrastructure for Exploration (LIFE) project, which aims
at building a multi-antenna radio telescope, using the LOIS technology,
on the far side of the moon.

\begin{figure}
\centering
 \begin{minipage}{\textwidth}
  \includegraphics[width=.49\textwidth]{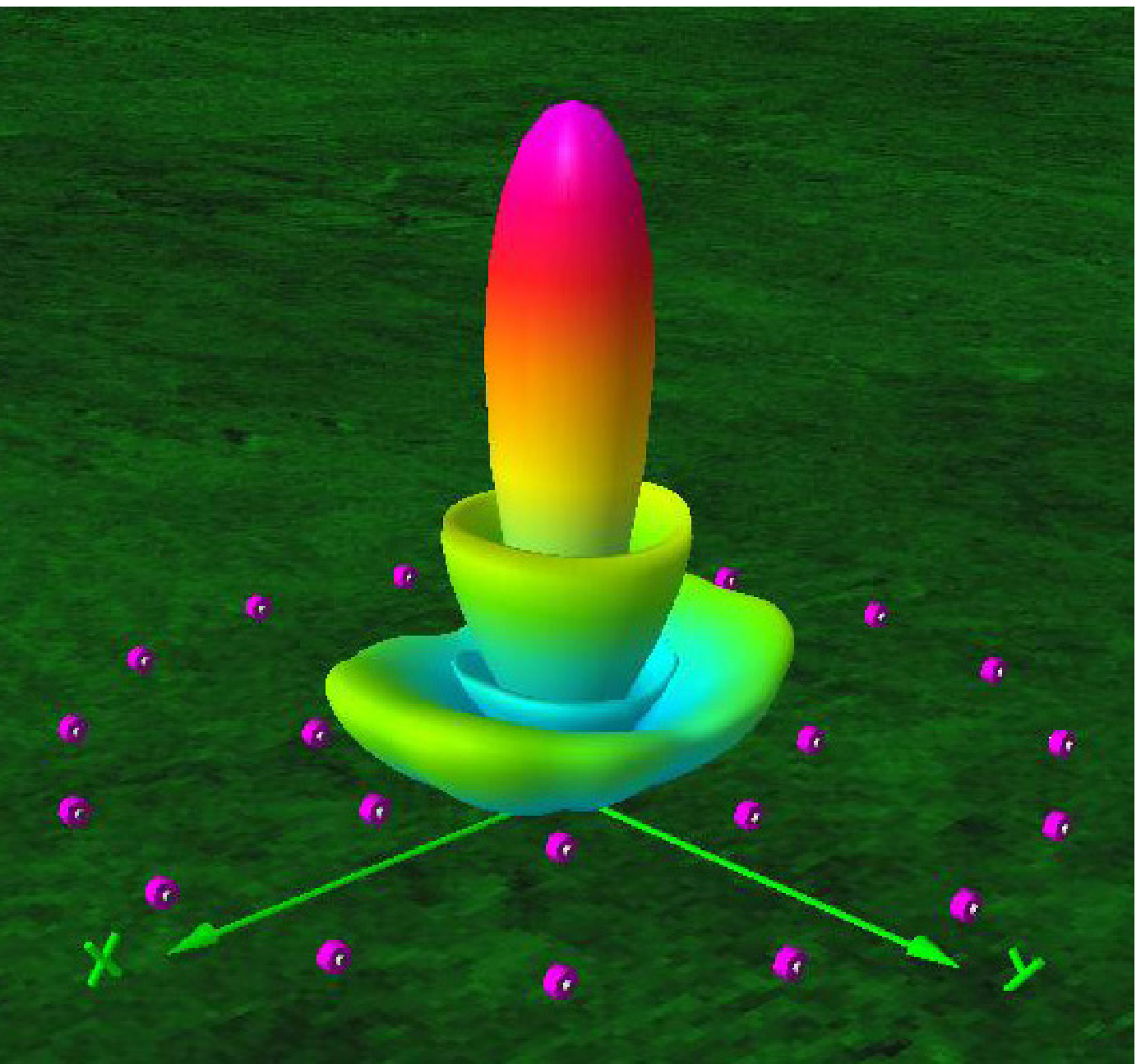}
  \hfill
  \includegraphics[width=.49\textwidth]{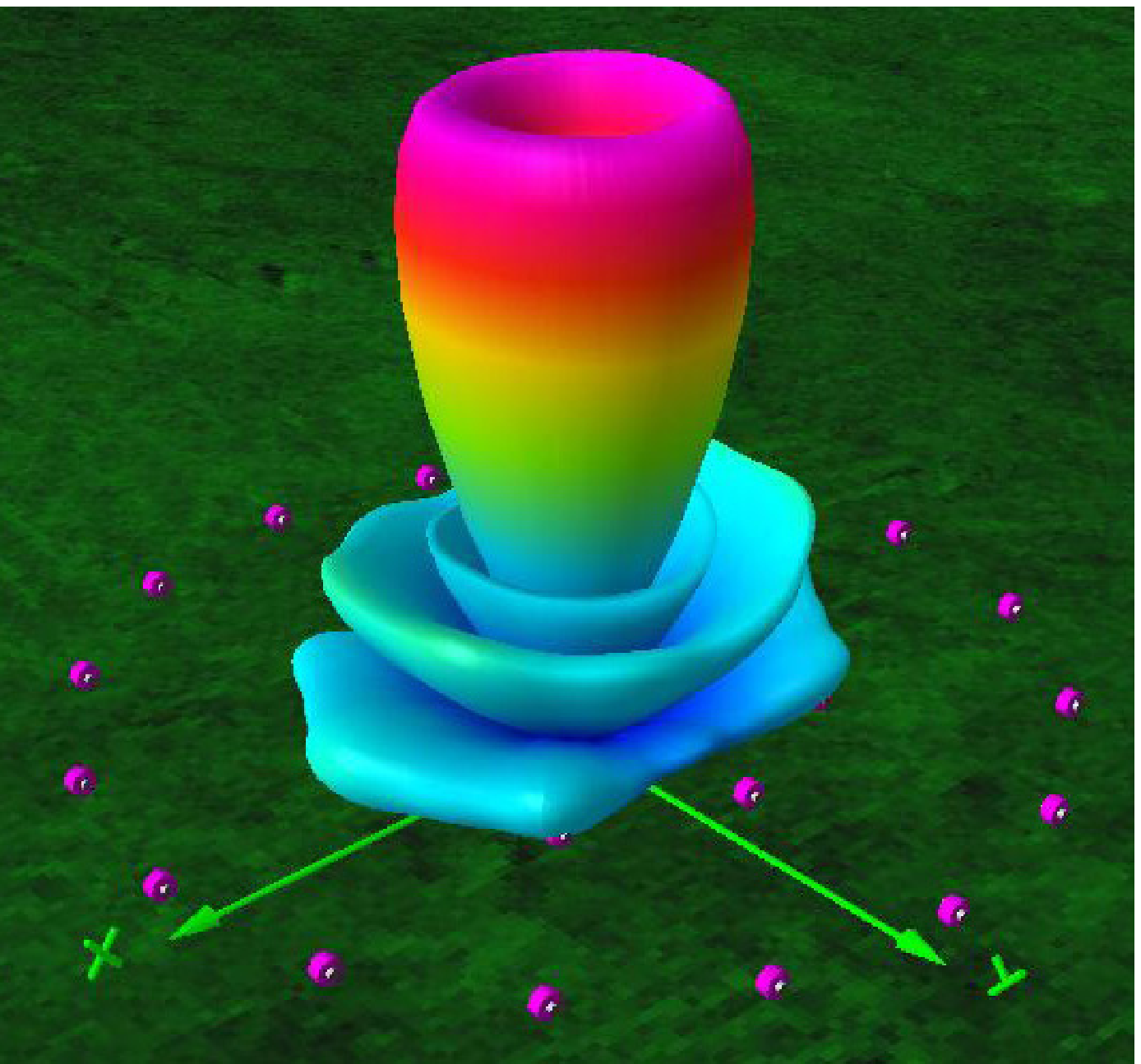}
  \\[1ex]
  \includegraphics[width=.49\textwidth]{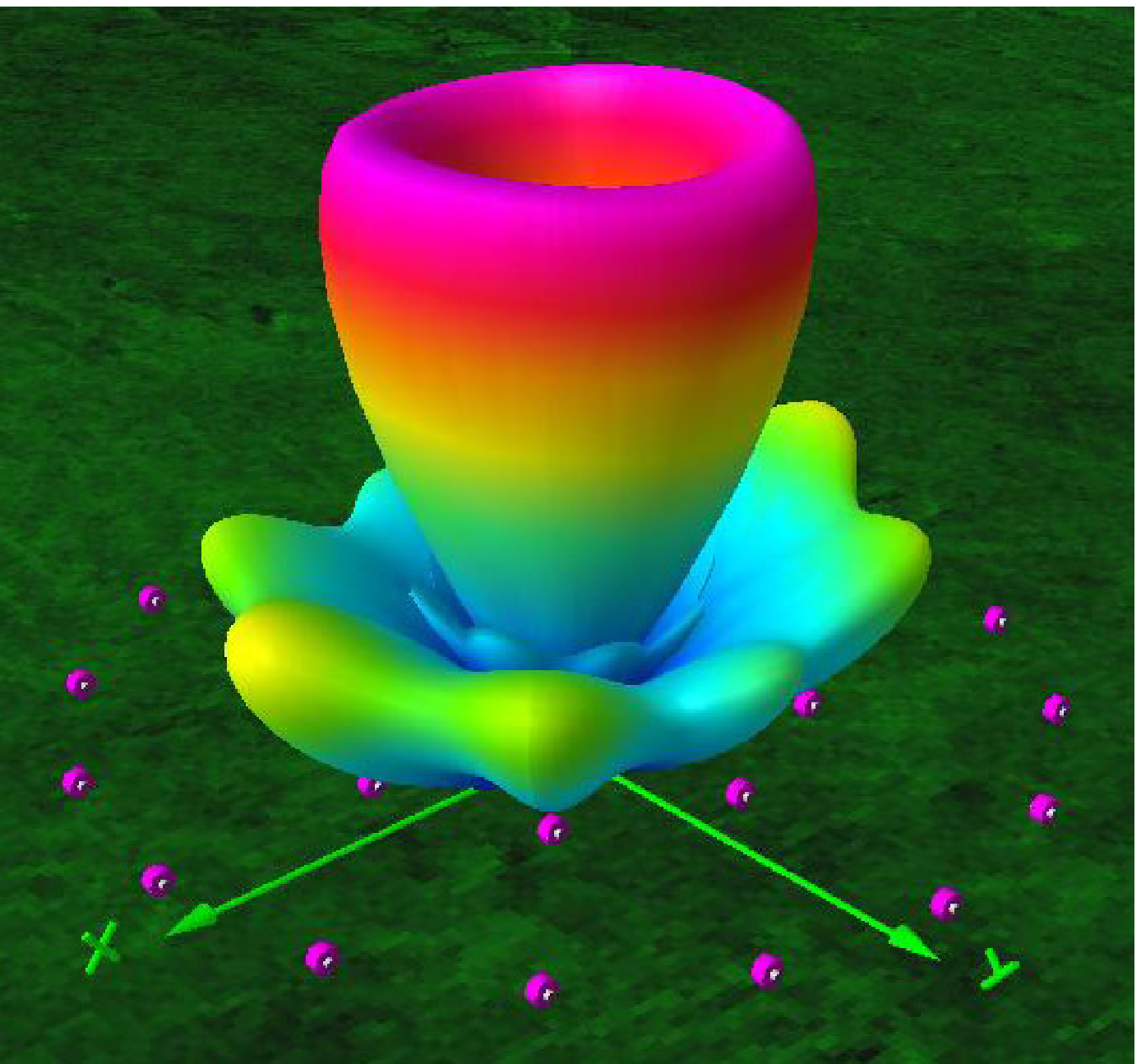}
  \hfill
  \includegraphics[width=.49\textwidth]{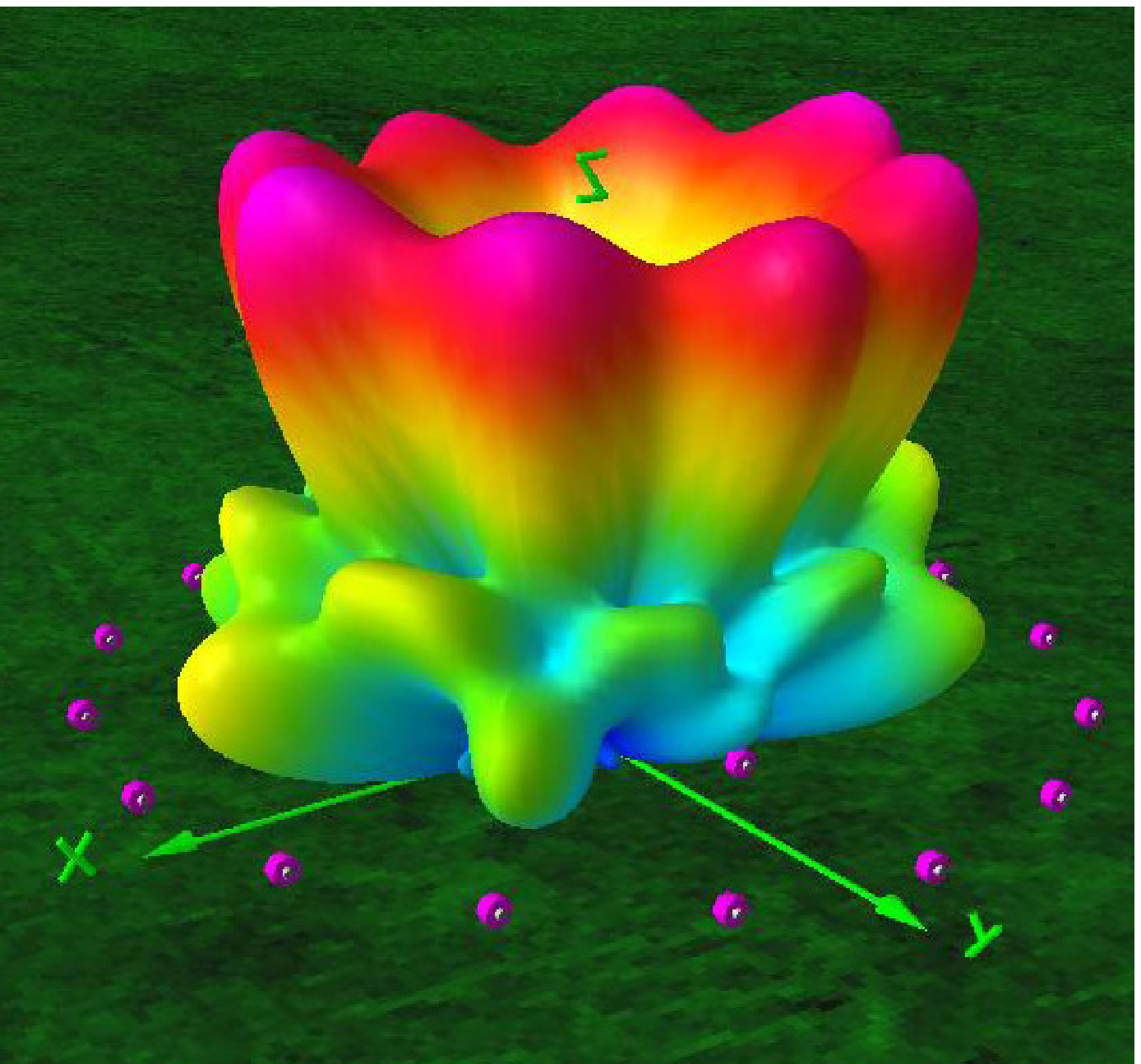}
 \end{minipage}
 \caption{%
 Radio beams generated by one ring of 8 antennas and radius $\lambda$
 plus a concentric ring of 16 antennas and radius $2\lambda$; all
 antennas are $0.25\lambda$ over ground.  These plots show the influence
 on the radiation pattern of (a) the OAM $l=0$ (upper left); (b) $l=1$
 (upper right); (c) $l=2$ (lower left); and (d) $l=4$ (lower right).
 }
\label{fig:patterns}
\end{figure}

The technique of applying OAM to radio beams opens for some very
interesting and powerful applications in astrophysics, space physics,
plasma physics and wireless communications \cite{Thide&al:PRL:2007}.
For instance, the antenna patterns produced can be useful because of
their basic shapes; see figure~\ref{fig:patterns}.  If, for example, we
want to probe the solar corona but not the Sun itself, the annular
intensity pattern of a beam carrying OAM is ideal since the intensity
minimum in the centre of the beam could be placed over the sun and the
rest of the beam over its corona.

\begin{figure}
\centering
 \includegraphics[width=1.\textwidth]{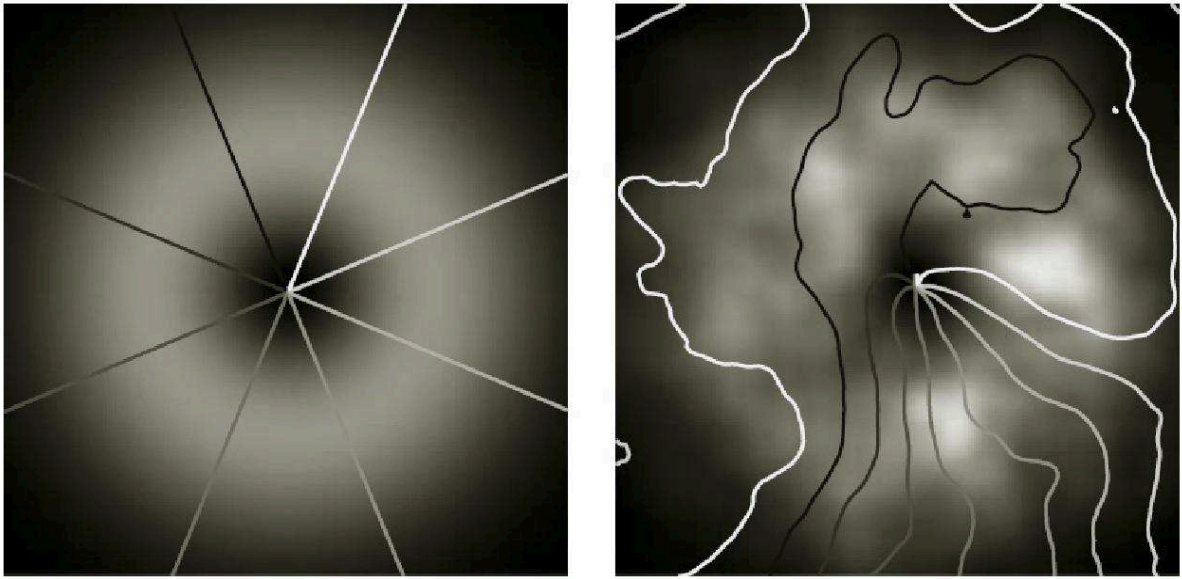}
 \caption{%
  Example intensity (gray scale) and phase map ($\pi/4$-spaced contours) of
  a pure OAM-carrying Laguerre-Gaussian beam (left) and the same beam but
  propagated through Kolmogorov turbulence (right). Adapted from
  \cite{Paterson:PRL:2005}. 
  }
\label{fig:turbulence}
\end{figure}

That the radio OAM can be a sensitive detector of turbulence in the
propagation medium (\eg, the turbulent ionospheric plasma for radio
astronomical signals) should be clear from figure~\ref{fig:turbulence}
which shows the result of a numerical simulation where a
Laguerre-Gaussian beam propagated through Kolmogorov turbulence
\cite{Paterson:PRL:2005}. The total angular momentum, \ie, the sum of
the spin angular momentum (wave polarisation) and OAM, for a given
volume of the plasma (including its vorticity) and a radio beam passing
through, or in any other way interacting with, this plasma volume is a
conserved quantity.  Vorticity in a medium is a clear signature of
turbulence in the same medium.  Hence, the OAM radio technique, which
measures the vorticity in radio signals, may be used to diagnose
plasma vorticity remotely.

%% file: acknowledgements.tex
\ack

The author wishes to thank his collaborators Holger Then, Johan
Sj\"oholm, Kristoffer Palmer, Jan Bergman, Tobia Carozzi, Yakov Istomin,
Nail Ibragimov, and Raisa Khimatova, without whose hard, dedicated
effort this paper could not have been written.  The financial support
from the Swedish Governmental Agency for Innovation Systems (VINNOVA) is
gratefully acknowledged.

%% file: references.tex
\section*{References}

\bibliographystyle{ieeetr}
\bibliography{eps}